\documentclass[preprintnumbers, sort&compress, floatfix]{revtex4}
\usepackage{amsmath}
\usepackage{amsfonts}
\usepackage{amssymb}
\usepackage{graphicx}
\usepackage{bm}
\usepackage{hyperref}
\def\be{\begin{equation}}
\def\ee{\end{equation}}
\def\bi{\bibitem}

\begin{document}
\title{Exact solutions of an anisotropic universe in a modified teleparallel gravity model via the Noether and B.N.S. approaches}
\author{‪Behzad Tajahmad$ $}
\email{behzadtajahmad@yahoo.com}
\affiliation{Faculty of Physics, University of Tabriz, Tabriz, Iran}
\begin{abstract}
In this paper, we present the Noether symmetries of locally rotationally symmetric Bianchi type I (LRS BI), an anisotropic model, in the context of the teleparallel gravity. We study a certain modified teleparallel theory based on the action that, in particular, contains a coupling between the scalar field and field strength (magnetism part). We derive the symmetry generators and show that, by means of cyclic variables approach, we can not obtain a suitable solution for field equations. Hence by the use of B.N.S. approach, we solve the equations which carry Noether currents as well. By data analysis of the obtained results, we show compatible results with observational data at the last half the age of universe which is accelerating.

\end{abstract}
\maketitle

\section{\bf{Introduction}}
Investigation of the essence and mechanism of the acceleration of our universe in the last decade, confirmed by some observational data on the account of supernova type IA (SNe Ia) \cite{1t1,1t2}, cosmic microwave background radiation (CMB)\cite{2t1, 2t2}, large-scale structure of the universe \cite{3t1, 3t2}, baryon acoustic oscillations \cite{4}, weak lensing \cite{5}, have lionized with cosmologists. There are two solutions for the elucidation for this problem; dark energy with negative pressure, understanding the nature of which, is one of the main problems in the research area, and the extended gravity (modified gravity) which has been a competitive alternative to the dark energy scenario. One of the routes to extend gravitational theories is Teleparallel Gravity (TG) where, the torsion scalar $T$ which consisting of the Weitzenb\"{o}ck connection, describes the action \cite{6t1, 6t2}. TG is demonstrably equivalent to general relativity \cite{7} and enables one to say that gravity is not due to curvature, but the torsion. $f(T)$ gravity, a generalized version of the so-called TG (proposed originally by Einstein \cite{8}), is easier to work with than $f(R)$ to analyze the cosmological evolution. This stems for the order of gravitational field equation which is two in the former one and four in the latter one. $f(T)$ gravity which provides alternative interpretations for the accelerating phases of the universe \cite{9t1, 9t2, 9t3}, has interesting cosmological solutions.

The symmetries of mechanical and physical systems have intimate connections with the associated conserved quantities. One of the methods for seeking the conserved quantities of the mechanical and physical systems is the Noether symmetry approach which is one of the most beautiful fruits of the calculus of variations. Noether theorem allows one to derive conserved quantities from the presence of variational symmetries \cite{10}. However, some hidden conserved currents may not be obtained by the Noether symmetry approach \cite{11,new}. Let a vector field $\textbf{X}$ can be defined on tangent space $TQ = (q, \dot{q})$ as
\begin{equation*}
\textbf{X} = \alpha^{\mu}(q) \frac{\partial }{\partial q^{\mu}} + \frac{d \alpha^{\mu}(q)}{dt} \frac{\partial }{\partial \dot{q}^{\mu}} ,
\end{equation*}
where $\alpha^{\mu}(q)$ are unknown functions on configuration space $Q=\{q\}$. Noether symmetry approach states that a function $F(q,\dot{q})$ is invariant under the transformation $\textbf{X}$ if
\begin{equation*}
\mathcal{L}_{\textbf{X}} F \equiv \alpha^{\mu}(q) \frac{\partial F}{\partial q^{\mu}} + \frac{d \alpha^{\mu}(q)}{dt} \frac{\partial F}{\partial \dot{q}^{\mu}}  = 0,
\end{equation*}
where $\mathcal{L}_{\textbf{X}} F$ is the Lie derivative of F. Specifically, if $\mathcal{L}_{\textbf{X}} L =0$, $\textbf{X}$ is a symmetry for the dynamics derived by $L$ (The Lagrangian). Therefore, it generates the following conserved quantity (constant of motion)
\be \label{I}
\mathbf{I} = \alpha^{i} \frac{\partial L}{\partial \dot{q}^{i}}.
\ee
Alternatively, utilizing the Cartan one-form
\be \label{teta}
\theta_{L} \equiv \frac{\partial L}{\partial \dot{q}^{i}} dq^{i}
\ee
and defining the inner derivative
\be \label{inner}
i_{\textbf{X}} \theta_{L} = \langle \theta_{L} , \textbf{X}\rangle
\ee
we get,
\be \label{cartan}
i_{\textbf{X}} \theta_{L} = \mathbf{I},
\ee
provided that \(\mathcal{L}_{\mathbf{X}}L=0\) holds. The Eq. (\ref{cartan}) is coordinate independent. The existence of Noether symmetry assures the presence of cyclic variables by a coordinate transformation such that the Lagrangian becomes cyclic in one of them. Using a point transformation, the vector field $\textbf{X}$ is rewritten as
\be \label{also lift}
\tilde{\textbf{X}} = \left(i_{\textbf{X}} dQ^{k} \right) \frac{\partial}{\partial Q^{k}} + \left[ \frac{d}{dQ^{k}} \left(i_{\textbf{X}} dQ^{k} \right) \right] \frac{\partial}{\partial \dot{Q}^{k}}.
\ee
If $\textbf{X}$ is a symmetry, so is $\tilde{\textbf{X}}$ (\textit{i.e.} $\tilde{\textbf{X}} L =0$), and a point transformation is chosen such that
\be \label{point equ}
i_{\textbf{X}} dQ^{1} = 1 , \qquad i_{\textbf{X}} dQ^{i} = 0 \qquad (i \neq 1).
\ee
It follows that
\be \label{tilde}
\tilde{\textbf{X}} = \frac{\partial}{\partial Q^{1}} , \qquad \frac{\partial L}{\partial Q^{1}} = 0,
\ee
therefore, $Q^{1}$ is a cyclic coordinate and the dynamics can be reduced. However, the change of coordinates is not unique and a clever choice would be advantageous \cite{b} and make the field equations to be easy solving. In the literature, applications of the Noether symmetry in generalized theories of gravity have been superabundantly studied (for example see \cite{c1, c2, c3, c4, c5, c6, c7, c8, c9, c10, c11, c12, c13, c14, c15, c16, c17, c18, c19, c20, c21, c22, c23, c24, c25, c26}).\\

On the other hand, there are cases that we have no solution with the Noether symmetry approach. Especially, the Noether approach, representing several conserved currents (Noether currents), is not conducive to any solution while matching all or a portion of them with field equations. The more currents there are the more problems pile up. On the other hand, hidden currents derivable from a continuity equation \cite{s1,s2}, are desirable to be included, but when doing so things get worse due to the abundance of currents. For solving the field equations, we have to remove some of the conserved currents. Moreover, symmetries have always played a central role in the conceptual discussion of classical and quantum physics.  The new approach, recently proposed by the author as ``B.N.S." (Beyond Noether Symmetry) approach may solve this problem \cite{12}. In such cases, the B.N.S.-approach would be useful and it paves this bumpy road. In almost all actions of extended gravity, we have some unknown functions (e.g. the scalar potential, the coupling functions with curvature and torsion and etc). Note that the standard way for defining the shape of these unknown functions is the Noether symmetry approach. But, the B.N.S.-approach tells that the main problem which excludes to obtain the analytical solution, is the form of these unknown functions, the main culprit in removing some of the conserved currents. In the case in which we have new forms of these unknown functions, then the problem can be solved. The B.N.S. approach carries it out in a simple way. Suppose that $F_{1} (\varphi)$, $F_{2} (\varphi)$, ..., $F_{n} (\varphi)$  are unknown functions where $\varphi = \varphi(t)$. First of all, we list all field equations and possible conserved currents, then use the maps as follows:\\
\begin{center}\label{B.T.Map}
\begin{tabular}{c c c c }
1. & $F_{1}(\varphi) \rightarrow F_{1}(t)$, & $ $ $ $ $ $ $\text{So we have:}$ $ $ $ $ & $F_{1}^{\prime}(\varphi) \rightarrow \frac{\dot{F_{1}}(t)}{\dot{\varphi}(t)}, $\\ \\
2. & $F_{2}(\varphi) \rightarrow F_{2}(t)$, & $ $ $ $ $ $ $ $ $ $ & $F_{2}^{\prime}(\varphi) \rightarrow \frac{\dot{F_{2}}(t)}{\dot{\varphi}(t)},$ \\ \\
$\vdots$ & $ $ $ $&$\Longrightarrow$  & $\vdots$ \\ \\
$n$. & $F_{n}(\varphi) \rightarrow F_{n}(t)$, & $ $ $ $ $ $ $ $ $ $ & $F_{n}^{\prime}(\varphi) \rightarrow \frac{\dot{F_{n}}(t)}{\dot{\varphi}(t)} $ $ ,$ \\ \\
\end{tabular}
\end{center}
where the prime indicates a derivative with respect to $\varphi$, and the dot indicates differentiation with respect to time. By substituting these in all equations, we may solve our ODE-system easily. After solving the system, we do an inverse map for obtaining the usual form of the unknown functions (i.e. depending upon $\varphi$). Perhaps, in some cases, the inverse map be hard to obtain. In such cases, one can do it numerically. In numerical inverse mapping, only two options are in order: requiring initial values or the time interval. Note that one can first carry out Noether approach for getting the conserved currents, and then proceed with this approach i.e. D.E-system = \{field equations + Noether conserved currents + other conserved currents such as hidden conserved currents\} without paying any attention to the form of the unknown functions which are obtained by the Noether approach. So, one could see that the form of the unknown function may be different from those derived from the Noether approach.\\

Generally speaking, the universe as confirmed by CMB temperature is anisotropic, but we premise it to be isotropic at large scales. One of the models that may describe such background is Bianchi type I model in which each direction has own scale factor. In this paper, we have studied a model in a special Bianchi type I which was named as locally rotationally symmetric Bianchi type I (LRS BI) in wich two of three spatial directions have the same scale factor.


\section{The model \label{II}}
The most generic action for a single field inflation,
\begin{equation*}
S = \int d^4x \sqrt{-g} \Bigg[{{M_{PL}^{2}}\over 2} R + {1\over 2}\phi_{,\mu}\phi^{,\mu} - V(\phi)- {1\over 4}{f^2(\phi)}F_{\mu
\nu}F^{\mu \nu} \Bigg],
\end{equation*}
in some articles such as Refs. \cite{gu,su2,su3,su4}, was investigated completely. These studies led to graceful results (inflation, late-time-accelerated expansion, etc.). Gauge fields are the main driving force for the inflationary background. There are several fields, such as the vector fields and the nonlinear electromagnetic fields, which are able to produce the negative pressure effects. In some papers such as Refs. \cite{su3,su4}, the authors used this model, perhaps, to answer the question whether or not this model may describe the late-time-accelerated expansion. Also, $T$-version of this action in FRW background has been recently studied \cite{12} and led to late-time-accelerated expansion as well as the strong matches of cosmological parameters with observational data (age of the universe, phase crossing, the present values of the scale factor, deceleration, Hubble, and state-finders parameters). Maybe, the main motivations for applying such models are the efforts made to obtain a unified model (with a single scalar field) which describes the stages of cosmic evolution. Now, we want to consider the $T$-version (teleparallel theory with $T$ ) of this action in an anisotropic background and finally compare our results with the Ref. \cite{15} which investigated the \(f(R)\)-version of this action in locally rotationally symmetric Bianchi type I. In this paper, our focus in data analysis of analytical solutions is on the last half the age of the universe which is accelerating.\\
We start with the following gravitational action \cite{12}
\be\label{action}
S = \int d^4x e \left[\frac{M^2_{Pl}}{2}T - \frac{1}{2} \varphi_{,\mu} \varphi^{,\mu} - V(\varphi) - \frac{1}{4} f^2(\varphi) F_{\mu\nu} F^{\mu\nu} \right]
\ee
where $e =$ det$(e^{j}_{\nu}) = \sqrt{-g}$ with $e^{j}_{\nu}$ being a vierbein (tetrad) basis, $T$ is the torsion scalar, $\varphi_{,\mu}$ stands for the components of the gradient of $\varphi$ which we assume it to be dependent upon time only and $V(\varphi)$ is the scalar potential. The vector potential \(\mathbf {A}\) of electromagnetic theory generates the electromagnetic field tensor via the geometric equation $\mathbf {F}$ = $-$(antisymmetric part of $\nabla {\mathbf {A}}$). Hence, for a given 4-potential \(A_{\mu }\), the field strength of the vector field is defined by \(F_{\mu \nu }= \partial_{\mu } A_{\nu } - \partial_{\nu } A_{\mu } \equiv A_{\nu , \mu } - A_{\mu , \nu }\). As is seen, in the action (\ref{action}) the gauge kinetic function \(f^2(\phi )\) is coupled to the strength tensor \(F_{\mu \nu }\).\\
The LRS BI universe model is given by
\be \label{metric}
ds^2 = dt^2 - a^2(t) dx^2 - b^2(t) \left[dy^2 + dz^2\right],
\ee
Here the metric potentials $a$ and $b$ are functions of time alone. According to this metric, without loss of generality, we introduce a homogeneous and anisotropic vector field as
\be
A_{\mu} = \left(A_{0}; A_{1},A_{2},A_{3}\right)= \left(\chi(t); 0, \frac{A(t)}{\sqrt{2}}, \frac{A(t)}{\sqrt{2}} \right)
\ee
whence we get
\be
F_{\mu\nu} F^{\mu\nu} = \frac{- 2 \dot{A}^2}{b^2}.
\ee
However, one can choose the gauge $A_{0} = \chi(t) = 0$, by using the gauge invariance \cite{gu}. Note that we have assumed the direction of the vector field does not change in time, for simplicity. The torsion scalar for the metric (\ref{metric}) can be found as
\be
T = - 2 \left(2 \frac{\dot{a}}{a}\frac{\dot{b}}{b} + \frac{\dot{b}^2}{b^2} \right)
\ee
in which the dot denotes a derivative with respect to time. The Lagrangian density corresponding to the action (\ref{action}) takes the form
\be \label{point-like}
L = -2 b \dot{b} \dot{a} - a \dot{b}^2 - \frac{1}{2} a b^2 \dot{\varphi}^2 - a b^2 V + \frac{1}{2} a f^2 \dot{A}^2
\ee
Here, we set the reduced Planck mass, \(M_{Pl}\), equal to 1. Regarding the point-like Lagrangian (\ref{point-like}), the Euler-Lagrange equations for the scale factors $a$ and $b$ become
\be \label{FE a}
2 \frac{\ddot{b}}{b}  + \frac{\dot{b}^2}{b^2} = \frac{\dot{\varphi}^2}{2} - \frac{f^2 \dot{A}^2}{ 2 b^2}+ V,
\ee
\be \label{FE b}
\frac{\ddot{a}}{a} + \frac{\ddot{b}}{b} + \frac{\dot{a}}{a} \frac{ \dot{b}}{b} = \frac{\dot{\varphi}^2}{2} + V.
\ee
For the vector $A$ and scalar field $\varphi$, the Euler-Lagrange equations take the following forms
\be \label{FE A}
\ddot{A} + \dot{A} \left(\frac{\dot{a}}{a} + 2 \frac{f^{\prime}}{f} \dot{\varphi} \right) = 0,
\ee
\be \label{FE phi}
\ddot{\varphi} + \dot{\varphi} \left( \frac{\dot{a}}{a} + 2 \frac{\dot{b}}{b}\right) = \frac{- f f^{\prime} \dot{A}^2}{b^2} + V^{\prime},
\ee
respectively, where the latter one is the Klein-Gordon equation and the prime indicates a derivative with respect to $\varphi$. And finally, the energy function,
\begin{equation*}
E_{L} = \sum_{j} \dot{q}^{j} \frac{\partial L}{\partial \dot{q}^{j}} - L \qquad ; \qquad q \in Q (\text{Configuration space}),
\end{equation*}
associated with the point-like lagrangian (\ref{point-like}) becomes
\be \label{FE E}
-2 \frac{\dot{a}}{a}\frac{\dot{b}}{b} - \frac{\dot{b}^2}{b^2} + \frac{1}{2} \dot{\varphi}^2 + \frac{f^2 \dot{A}^2}{2 b^2} + V = 0.
\ee
\section{Exact solutions via the Noether symmetry and B.N.S. approaches}
In this section, we want to use Noether symmetry approach for solving Eqs. (\ref{FE a})-(\ref{FE E}). The configuration space and tangent space for the point-like lagrangian (\ref{point-like}) are $Q = \{a, b, \varphi, A\}$ and $TQ = \{a, \dot{a}, b, \dot{b}, \varphi, \dot{\varphi}, A, \dot{A}\}$, respectively. Hence the infinitesimal generator of the Noether symmetry is
\be \label{generator}
\textbf{X} = \xi \frac{\partial}{\partial a} + \alpha \frac{\partial}{\partial b} + \beta \frac{\partial}{\partial \varphi} + \gamma \frac{\partial}{\partial A} + \xi_{,t} \frac{\partial}{\partial \dot{a}} + \alpha_{,t} \frac{\partial}{\partial \dot{b}} + \beta_{,t} \frac{\partial}{\partial \dot{\varphi}} + \gamma_{,t} \frac{\partial}{\partial \dot{A}},
\ee
where
\be
y = y (a, b, \varphi, A) \quad \longrightarrow \quad y_{,t} = \dot{a} \frac{\partial y}{\partial a} + \dot{b} \frac{\partial y}{\partial b} + \dot{\varphi} \frac{\partial y}{\partial \varphi} + \dot{A} \frac{\partial y}{\partial A} \quad ; \quad y \in \{\xi, \alpha, \beta, \gamma\}.
\ee
The existence of Noether symmetry implies the existence of a vector field \textbf{X} such that the Lie derivative of the Lagrangian with respect to the vector field vanishes, i.e.,
\be \label{condition.}
\mathcal{L}_{\textbf{X}} L = 0 \quad \longrightarrow \quad \xi \frac{\partial L}{\partial a} + \alpha \frac{\partial L}{\partial b} + \beta \frac{\partial L}{\partial \varphi} + \gamma \frac{\partial L}{\partial A} + \xi_{,t} \frac{\partial L}{\partial \dot{a}} + \alpha_{,t} \frac{\partial L}{\partial \dot{b}} + \beta_{,t} \frac{\partial L}{\partial \dot{\varphi}} + \gamma_{,t} \frac{\partial L}{\partial \dot{A}} = 0.
\ee
This yields an expression which is of second degree in $a$, $b$, $\varphi$, and $A$ and whose coefficients are functions of $a$, $b$, $\varphi$, and $A$, only. Thus to satisfy equation (\ref{condition.}), we obtain the following set of equations
\be \begin{split}
&\frac{\partial \alpha}{\partial a} = 0,\\ &\xi + 2 a \frac{\partial \alpha}{\partial b} + 2 b \frac{\partial \xi}{\partial b} =0,\\ &\xi b+2ab\frac{\partial \beta}{\partial \varphi}+2a\alpha=0,\\ &\xi f+2a \beta f^{\prime}+2af \frac{\partial \gamma}{\partial A} =0,\\
&\alpha+b \left(\frac{\partial \xi}{\partial a}+\frac{\partial \alpha}{\partial b} \right)=0,\\ &ab \frac{\partial \beta}{\partial a}+2\frac{\partial \alpha}{\partial \varphi}=0,\\ &ab^2 \frac{\partial \beta}{\partial b}+2a \frac{\partial \alpha}{\partial \varphi}+2b \frac{\partial \xi}{\partial \varphi}=0,\\ &-b^2 \frac{\partial \beta}{\partial A}+f^2 \frac{\partial \gamma}{\partial \varphi}=0,\\
&af^2 \frac{\partial \gamma}{\partial a} - 2b \frac{\partial \alpha}{\partial A}=0,\\ &af^2 \frac{\partial \gamma}{\partial b}-2a \frac{\partial \alpha}{\partial A}-2b \frac{\partial \xi}{\partial A}=0,\\ &\left(2a\alpha +\xi b \right)V+ba\beta V^{\prime}=0.
\end{split}\ee
Solving this system of partial differential equations leads to the following solutions
\be\label{sol 1}\begin{split}
\xi=-2c_{1}a, \quad \quad \alpha = c_{1}b, \quad& \quad \beta=0, \quad \quad \gamma=c_{1}A + c_{2},\\ f= \text{Any arbitrary function of $\varphi$} \quad &, \quad V=\text{Any arbitrary function of $\varphi$}.
\end{split}\ee
As we observe, Noether approach can not give the forms of unknown functions in this case.
Symmetry generators, $\textbf{X}_{i}$, on tangent space turn out to be
\be\label{X1}
\textbf{X}_{1}=-2a\frac{\partial}{\partial a}+b\frac{\partial}{\partial b}+A\frac{\partial}{\partial A},
\ee
\be\label{X2}
\textbf{X}_{2}=\frac{\partial}{\partial A}.
\ee
According to Eq. (\ref{I}), the following functions are the constants of motion
\begin{equation} \label{cons.1}
\textbf{I}_{1} = 4ab\dot{b}-2b \left(b\dot{a}+a\dot{b}\right)+Aaf^2 \dot{A} \equiv c_{1},
\end{equation}
\begin{equation}\label{cons.2}
\textbf{I}_{2} = af^2 \dot{A} \equiv c_{2},
\end{equation}
where $c_{1}$ and $c_{2}$ are arbitrary constants. We add (\ref{cons.1}) and (\ref{cons.2}) to the set of Eqs. (\ref{FE a})-(\ref{FE E}) and solve this system. We used cyclic variables approach for solving, but it produced a constant vector field, $A(t)= const.$, which does not make sense for the action (\ref{action}) because constant $A(t)$ removes the last term in the action (\ref{action}). This is compelling to use B.N.S. approach where without any loss of generality, we use the maps as follows
\be\label{map 1} \begin{split}\left\{
\begin{array}{ll}
i:& \hbox{$f(\varphi(t)) \longrightarrow f(t),$} \\ \\
ii:& \hbox{$V(\varphi(t)) \longrightarrow V (t).$}
\end{array}
\right. \end{split} \ee
Therefore, we have
\be \label{map 2} \begin{split}\left\{
\begin{array}{ll}
i:& \hbox{$f^{\prime}(\varphi)\longrightarrow \frac{\dot{f}(t)}{\dot{\varphi}(t)},$} \\ \\
ii:& \hbox{$V^{\prime}(\varphi)\longrightarrow \frac{\dot{V}(t)}{\dot{\varphi}(t)}.$}
\end{array}
\right. \end{split} \ee

\begin{figure}
\centering
\includegraphics[width=7 in, height=3 in]{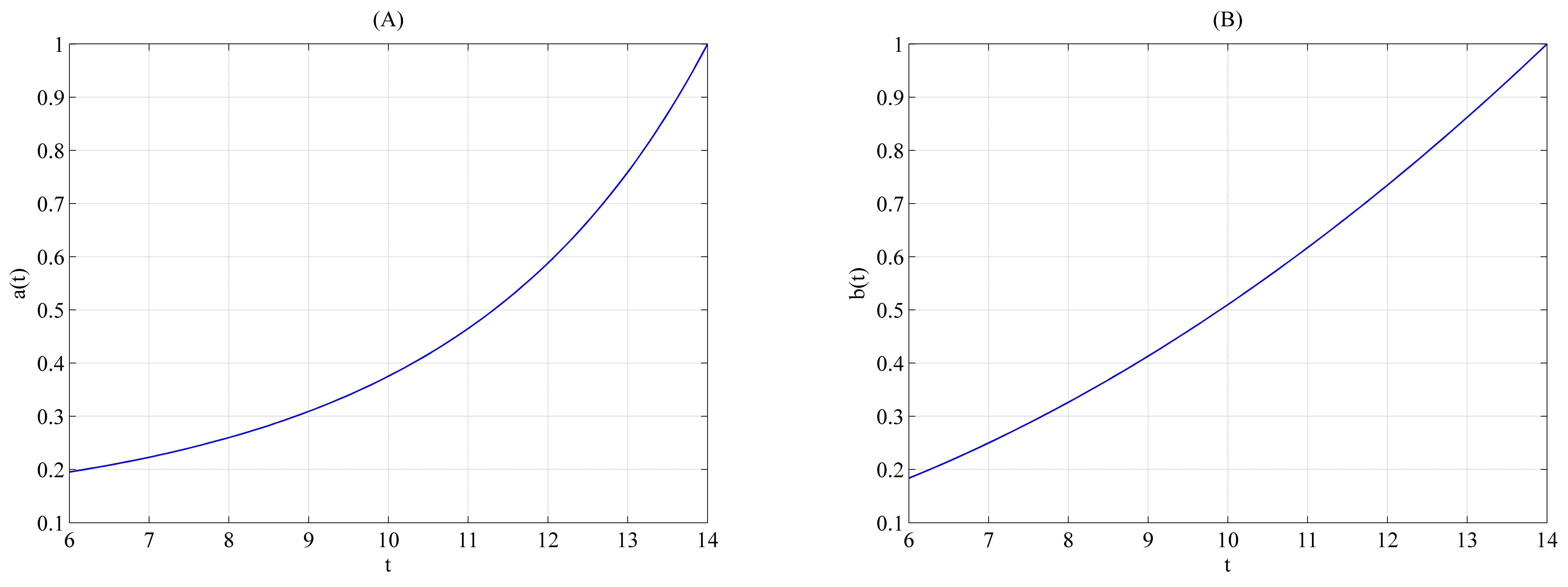}\\
\caption{Plots (A) and (B) indicate the scale factors $a(t)$ and $b(t)$ versus time $t$ respectively at the time range [6 , 14].}\label{fig1}
\end{figure}

Substituting Eqs. (\ref{map 1}) and (\ref{map 2}) in Eqs. (\ref{FE a})-(\ref{FE E}) and (\ref{cons.1})-(\ref{cons.2}), we obtain a modified system of differential equations. One set of solution for this system is
\be \begin{split} \label{sol FE}
a(t) &= a_{0} e^{k_{1} t^2}, \quad \quad b(t)=b_{0} t^2, \quad \quad
A(t)= \frac{4 k_{1} a_{0} b_{0}^2 t^3 e^{k_{1} t^2} \left(k_{1} t^2 -1 \right) +c_{1}}{c_{2} k_{1}}, \quad \quad \\ \\
f(t) &= \frac{c_{2} e^{- k_{1} t^2}}{2 a_{0} b_{0} t \sqrt{2t^4 k_{1}^2+3k_{1}t^2-3}}, \quad \quad V(t)=\frac{2 (2k_{1}t^2+3)}{t^2},\\ \\
\varphi(t)&=\int \left( \frac{2\sqrt{2t^4 k_{1}^2+k_{1} t^2-2}}{t} \right)dt
\\&=\sqrt{2t^4k_{1}^2+t^2 k_{1}-2}+\frac{\ln \left(\sqrt{2}k_{1}t^2+\sqrt{2t^4 k_{1}^2+t^2 k_{1}-2}+2^{-3/2}\right)}{2\sqrt{2}}-\sqrt{2} \arctan \left(\frac{t^2 k_{1}-4}{2\sqrt{2(2t^4 k_{1}^2 + t^2 k_{1}-2)}} \right).
\end{split}\ee
Limpidly, doing an inverse map for reaching at the usual form of the unknown functions (i.e. depending upon $\varphi$) in this case is challenging, so we must do it numerically. Maybe, it appears that the shape of the unknown functions (i.e. $f$ and $V$) in terms of the scalar field will be complicated due to the form of the scalar field $\varphi$. But, note that we can not state this, because they are unknown in terms of the scalar field, hence we do it numerically. So we are not able to do such discussions.\\
Hamiltonian constraints, $E_{L}=0$ (Eq. \ref{FE E}), $\textbf{I}_{1} = c_{1}$ (Eq. \ref{cons.1}), and $\textbf{I}_{2} = c_{2}$ (Eq. \ref{cons.2}) are three conserved currents carried by this set of solution. They correspond to $\textbf{X}=\partial/\partial t$, $\textbf{X}_{1}$ (Eq. (\ref{X1})), and $\textbf{X}_{2}$ (Eq. (\ref{X2}))respectively.\\

\begin{itemize}
  \item \textbf{Satisfaction of Maxwell's equations}
\end{itemize}
Here, we want to answer the question whether, with the obtained form of the vector potential, the Maxwell's equations are satisfied. For this purpose, we must utilize Maxwell's equations in curved spacetime, which in terms of the components of the field tensor $\mathbf{F}$ are \cite{gravitation}
\begin{equation}\label{M1}
F_{\alpha \beta , \gamma}+F_{\beta \gamma , \alpha} + F_{\gamma \alpha , \beta} =0,
\end{equation}
\begin{equation}\begin{split}\label{M2}
{F^{\alpha \beta}}_{,\beta} = -4 \pi J^{\alpha}; \qquad
\text{such that,} \qquad
\left\{
\begin{array}{ll}\begin{split}
\text{if $\alpha = 0$}: \quad& \hbox{$J^0 = \rho = \text{charge density,}$} \\ \\
\text{if $\alpha \neq 0$}:\quad& \hbox{$(J^1, J^2, J^3) = \text{components of current density,} $}
\end{split}\end{array}
\right. \end{split}
\end{equation}
where $\{J^{\alpha}$ ; $\alpha \in \{0, 1, 2, 3\}\}$ are the components of the 4-current $\mathbf{J}$. In a nutshell, through Eq. (\ref{M1}) magnetodynamics and magnetostatics, and through Eq. (\ref{M2}) electrodynamics and electrostatics are unified in one geometric law. The usual form of Maxwell's equations may be reached at easily since Eq. (\ref{M1}) reduces to $\mathbf{\nabla} \mathbf{\cdot} \mathbf{B} = 0$ when one takes $\alpha = 1$, $\beta = 2$, $\gamma = 3$; and it reduces to $\partial \mathbf{B} / \partial t + \mathbf{\nabla} \times \mathbf{E} = 0$ when one sets any index, e.g., $\alpha = 0$, and finally, with Eq. (\ref{M2}) two of Maxwell's equations, $\mathbf{\nabla \cdot E }= 4 \pi \rho$ (the electrostatic equation), $\partial \mathbf{E} / \partial t - \mathbf{\nabla} \times \mathbf{B} = -4 \pi \mathbf{J}$ (the electrodynamic equation), are obtained by putting $\alpha = 0$ and $\alpha \neq 0$, respectively. \\

For the electromagnetism part of the action (\ref{action}), i.e.
\begin{equation}\label{M3}
\mathcal{L}_{EM} = -\frac{1}{4}\int d^4 x \sqrt{-g}f^2(\varphi) F_{\mu \nu}F^{\mu \nu} = -\frac{1}{4}\int d^4 x \sqrt{-g} g^{\alpha \beta} g^{\mu \nu} f^2(\varphi) F_{\mu \alpha} F_{\nu \beta},
\end{equation}
Eqs. (\ref{M1}) and (\ref{M2}) read
\begin{equation}\label{M41}
\partial^{\alpha} \left(\sqrt{-g} f^2(\varphi) F^{\beta \gamma} \right) + \partial^{\beta} \left(\sqrt{-g} f^2(\varphi) F^{\gamma \alpha} \right) +\partial^{\gamma} \left(\sqrt{-g} f^2(\varphi) F^{\alpha \beta} \right) =0,
\end{equation}
\begin{equation}\label{M42}
\partial _{\mu} \left[\sqrt{-g} g^{\alpha \beta} g^{\mu \nu} f^2(\varphi) F_{\nu \beta}\right] =0  \quad \longrightarrow \quad \left(\sqrt{-g}f^2(\varphi) F^{\alpha \mu} \right)_{,\mu} = 0,
\end{equation}
respectively. Note that in our studying case, we have $\mathbf{J}=(J^0, J^1, J^2, J^3) = (0, 0, 0, 0)$. After simplifying, both equations (\ref{M41}, \ref{M42}) lead to the same equation, viz,
\begin{equation}\label{Mo}
\frac{\partial}{\partial t} \left(a f^2 \dot{A} \right) = 0.
\end{equation}
Clearly, this equation is equivalent to the third field equation (i.e. Eq.(\ref{FE A})). Hence, Eqs. (\ref{M1}) and (\ref{M2}) have been satisfied automatically when the solution for the field equations was found. Therefore, the results are consistent with all Maxwell's field equations.\\

\section{Data analysis of the exact solution}
First of all, let us give a brief review of some of the cosmological parameters in our anisotropic background (\ref{metric}).
\subsection{Some Definitions.}
\begin{description}
  \item[Hubble Parameter.] This parameter may be used for extracting the age of the universe and showing the expansion of our universe. Mathematically, the expansion of the universe in the FRW background is described by a scale factor $a(t)$, which can be interpreted as the size of the universe at a time $t$, but relative to some reference size (typically chosen to be the current size). The Hubble parameter for FRW background, is defined as $H_{FRW} = \dot{a}(t)/a(t)$. In the other backgrounds such as LRS Bianchi-I (\ref{metric}), we must introduce an average scale factor. In our case, we can define it as geometrical average, i.e. $a_{ave.}=(ab^2)^{1/3}$. So, for LRS BI, the mean Hubble parameter is given by
\begin{equation*}
  H=\frac{\dot{a}_{ave.}}{a_{ave.}} =\frac{1}{3} \left(\frac{\dot{a}}{a}+\frac{2\dot{b}}{b}\right).
\end{equation*}
  \item[Deceleration Parameter.] The deceleration parameter is
\begin{equation*}
  q = \frac{d}{dt} \left(\frac{1}{H}\right) - 1,
\end{equation*}
which acts as a tool to determine the nature of cosmic expansion; i.e. for $q>0$ and $q<0$ we have decelerated and accelerated expansion respectively. Therefore, in this case (LRS BI), it takes the following form
\begin{equation*}
  q= - \frac{3\ddot{a}ab^2+6\ddot{b}a^2b-2\dot{a}^2b^2+4\dot{a}\dot{b}ab-2\dot{b}^2a^2}{\left(\dot{a}b+2\dot{b}a \right)^2}.
\end{equation*}
  \item[State Finders.] The $(r,s)$ parameters are the state finders that help to check the correspondence of constructed models with the standard universe models. The value of $(r,s)=(1,0)$ represents correspondence of the constructed model with standard $\Lambda$CDM universe model. These are defined in terms of the deceleration and Hubble parameters as \cite{14}
\begin{equation*}
r=q+2q^2-\frac{\dot{q}}{H} \qquad , \qquad s=\frac{r-1}{3 \left(q-\frac{1}{2}\right)}.
\end{equation*}
  \item[Om-diagnostic.] The \textit{Om-diagnostic} is an important geometrical diagnostic proposed by Sahni \textit{et al.} \cite{17}, in order to classify the different dark energy (DE) models. The $Om$ is able to distinguish dynamical DE from the cosmological constant in a robust manner both with and without reference to the value of the matter density. The $Om$ analysis has been applied to several models (for example, see \cite{18t1}). It is defined as
\begin{equation}\label{Om}
Om(z) \equiv \frac{\left[\frac{H(z)}{H_{0}}\right]^2 -1}{\left(1+z \right)^3 -1},
\end{equation}
where $z$ and $H_{0}$ are the redshift ($z=a^{-1}-1$) and the present value of Hubble parameter, respectively. For dark energy with a constant equation of state (EoS) $\omega$, it reads
\begin{equation}\label{Om-w}
Om(z) = \Omega_{m0} + (1-\Omega_{m0}) \frac{(1+z)^{3(1+\omega)}-1}{(1+z)^{3}-1},
\end{equation}
so, $Om(z) = \Omega_{m0}$ states the $\Lambda$CDM model, therefore the regions $Om(z) > \Omega_{m0} \simeq 0.3$ and $Om(z) < \Omega_{m0} \simeq 0.3$ correspond with quintessence ($\omega > -1$) and phantom ($\omega < -1$), respectively. In our anisotropic metric, the redshift $z$ is given by
\be
z=\frac{1}{a_{ave.}} -1.
\ee
\end{description}

\subsection{Data Analysis.}
We present five figures to demonstrate the results obtained through the following choices for constants of integrations
\begin{equation}\label{selections for constants}
a_{0}=e^{-2}, \quad \quad b_{0}=\frac{1}{196}, \quad \quad k_{1}=\frac{1}{98}, \quad \quad c_{1}=0, \quad \quad c_{2}=0.5,
\end{equation}
\begin{figure}
\centering
\includegraphics[width=7 in, height=2.7 in]{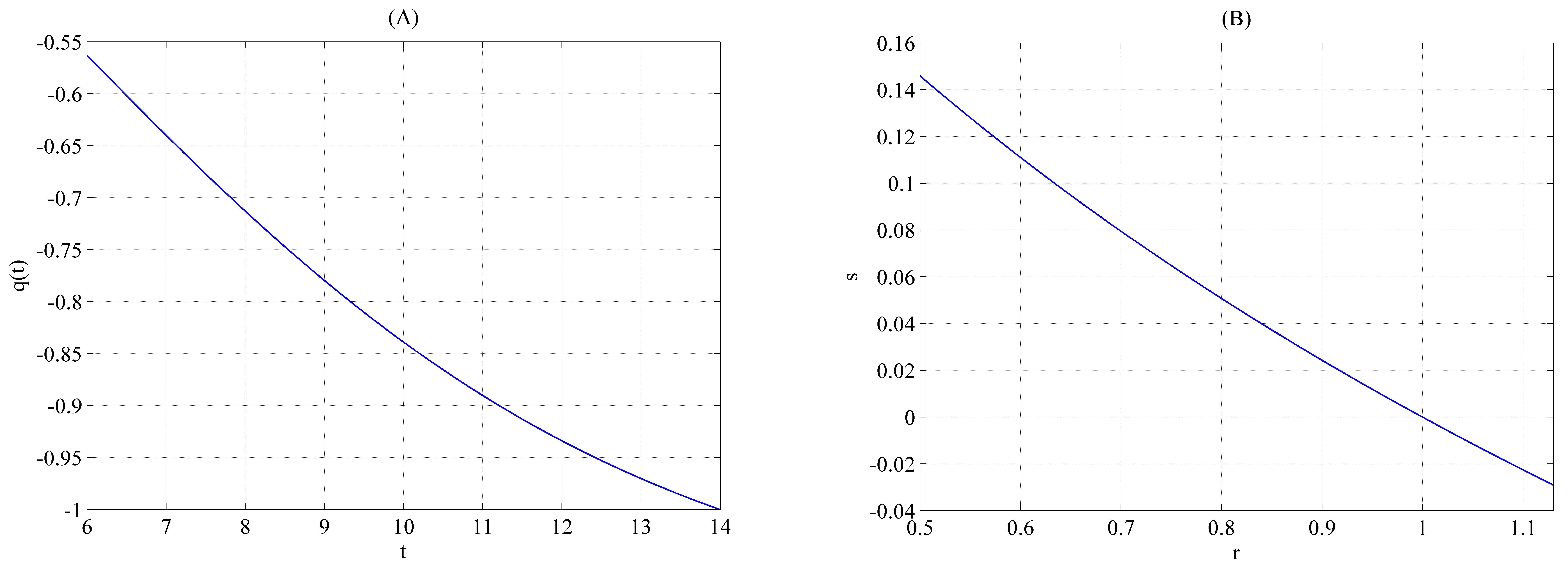}\\
\caption{Plots (A) indicates the deceleration parameter $q(t)$ versus time $t$ at the time range [6 , 14] and plot (B) shows $(r,s)$ parameters (state finders).}\label{fig2}
\end{figure}
\begin{figure}
\centering
\includegraphics[width=3.5 in, height=2.7 in]{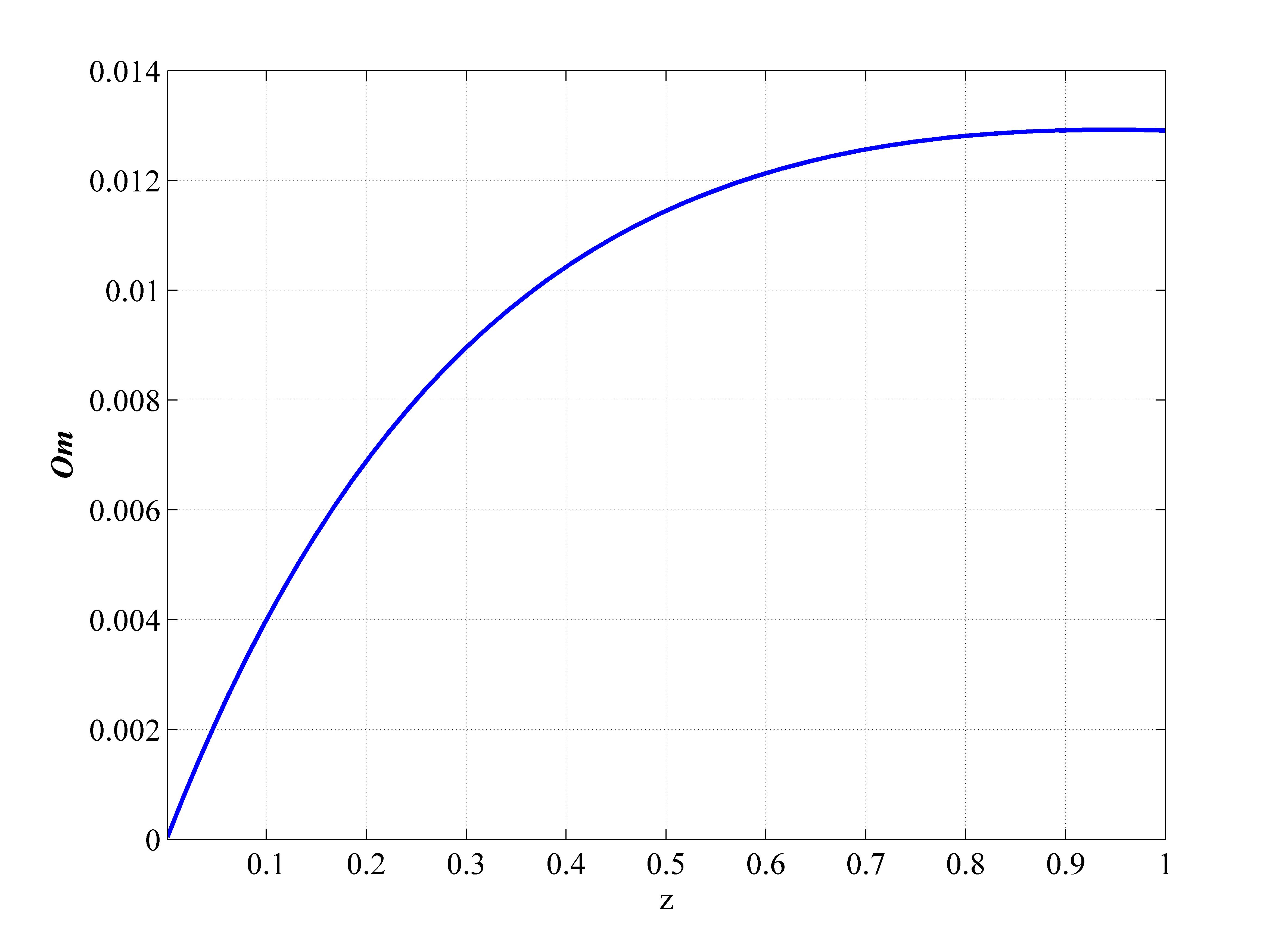}\\
\caption{Plot indicates the $Om$ diagnostic versus redshift $z$ at the redshift range [0.001,1].}\label{fig5}
\end{figure}
and discuss the cosmological behavior of LRS BI universe model. We use the time unit $1 Gyr \equiv \mathbf{1}$ in plotting.

As mentioned above, our focus in data analysis of analytical solutions is on the last half the age of the universe which is accelerating. Hence, we present plots at the time range $[6, 14]$.\\
With the above selections, Hubble, deceleration, and state finder parameters read
\begin{equation*}\begin{split}
H&=\frac{t}{147}+\frac{4}{3t}, \quad \quad q= - \frac{t^4+539t^2+9604}{\left(t^2+196\right)^2}, \quad \quad \\ \\
r&=\frac{t^6+1029t^4+115248t^2-941192}{(t^2+196)^3}, \quad\quad s=\frac{49(19208-t^4)}{(245t^2+9604+0.5t^4)(t^2+196)}.
\end{split}\end{equation*}

\textit{\textbf{Interpretations of the results:}} Figures \ref{fig1}(A) and \ref{fig1}(B) indicate the scale factors with increasing nature expressing the expansion of the universe. However as is clear in (\ref{sol FE}), $a(t)$ grows faster than $b(t)$ because of $a \propto \exp(b)$. Obviously, the expansions in the directions are accelerated, as the positive gradients of both curves are growing with time. The present amounts of both scale factors are one, as we expected.
\begin{figure}
\centering
\includegraphics[width=7 in, height=2.7 in]{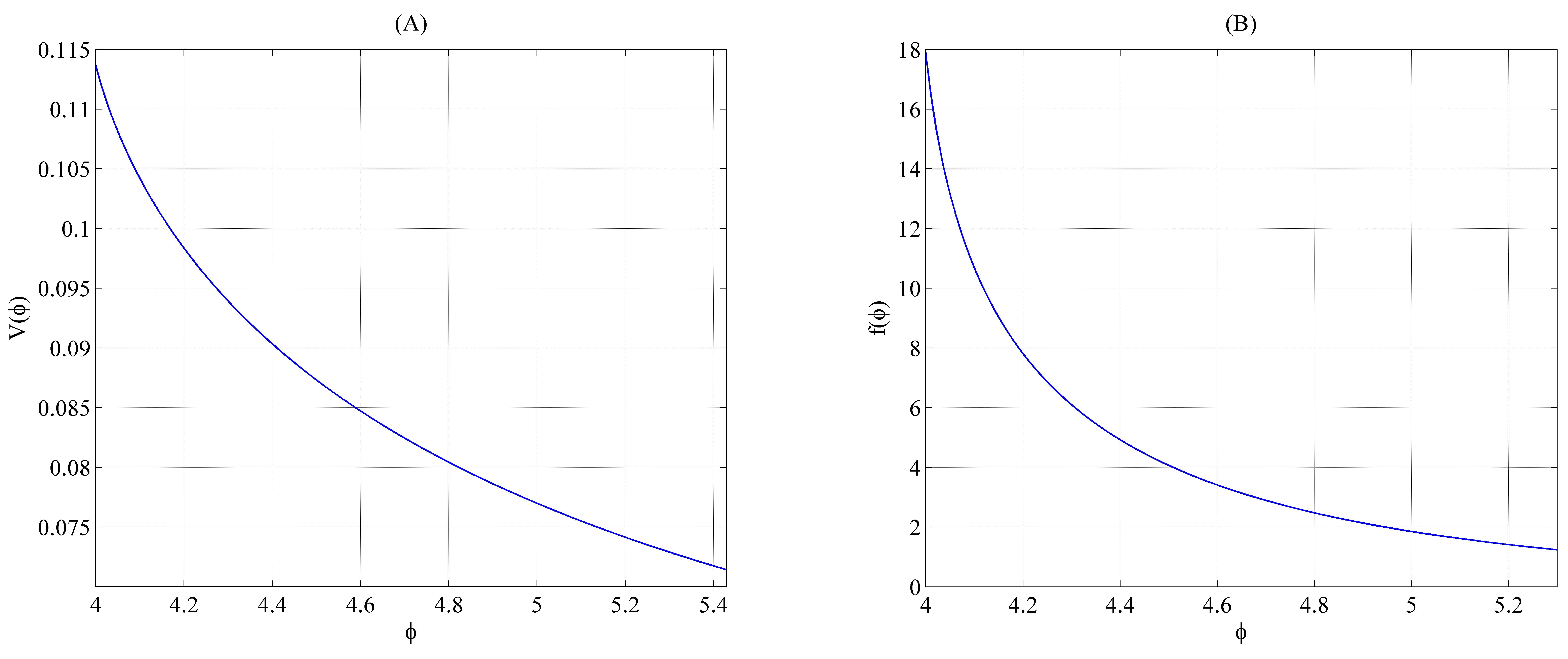}\\
\caption{Plots (A) and (B) indicate the scalar potential $V(\varphi)$ and coupling function $f(\varphi)$ versus scalar field $\varphi$ respectively.}\label{fig3}
\end{figure}
\begin{figure}
\centering
\includegraphics[width=7 in, height=2.7 in]{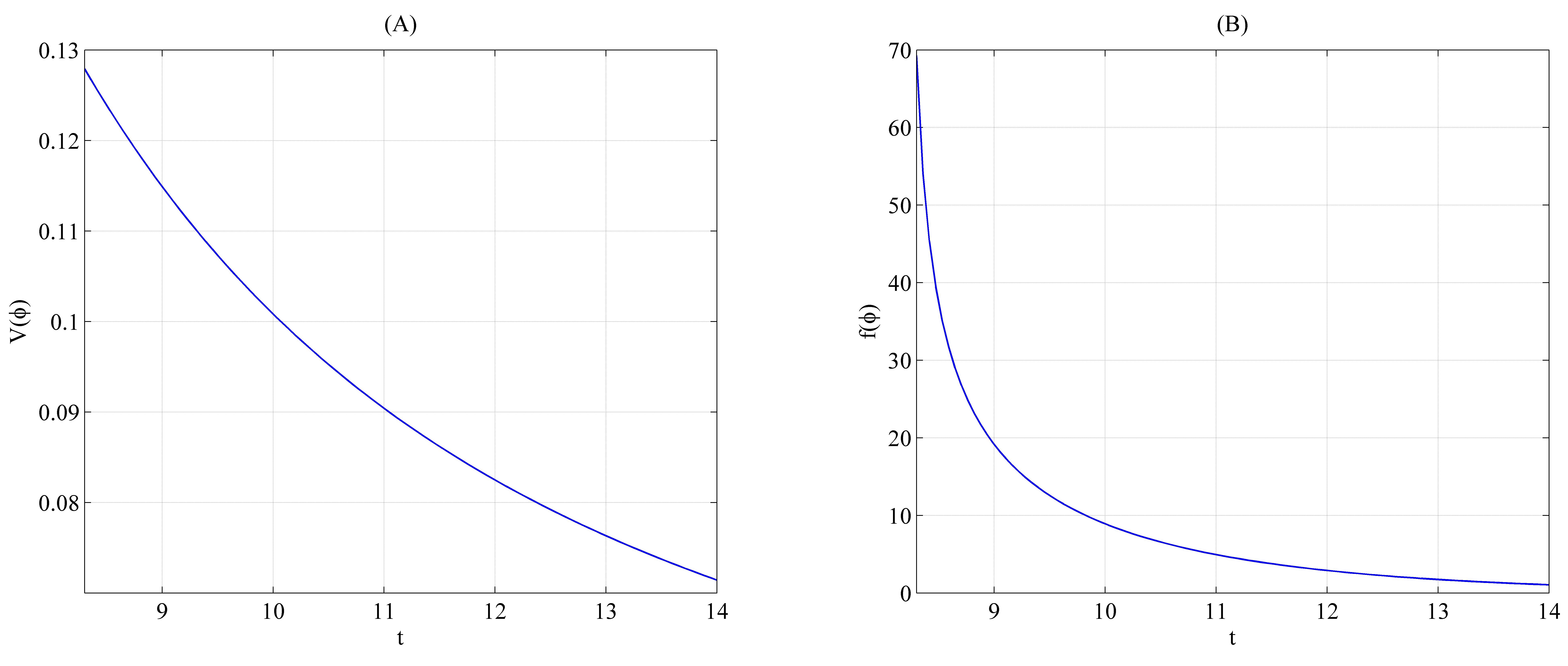}\\
\caption{Plots (A) and (B) indicate the scalar potential $V(\varphi)$ and coupling function $f(\varphi)$ versus time $t$ respectively.}\label{fig4}
\end{figure}
Figure \ref{fig2} (A) indicates the negative behavior of deceleration parameter implying an accelerating universe at the last half age of the universe. Also, it is close to $q_{0} = -1$ at present time. As we know, it coincides with observational data, as they tell $q_{0}=1$. A trajectory passes through a point $(r,s) = (1,0)$, which indicates a close correspondence with the $\Lambda$CDM universe model, has been showed with Figure \ref{fig2} (B). According to the Figure \ref{fig5}, we learn that at the half the age of the universe, our universe is in phantom phase because of $Om(z) < \Omega_{m0} \sim 0.3$. Observational data represent that we had a phase crossing from quintessence to phantom phase and it occurred at about the half the age of the universe, hence the present phase of the our universe is phantom. So, the resulting model has a perfect agreement with this point.  Figure \ref{fig3} (A) presents a subtractive behavior of scalar potential $V(\varphi)$ versus scalar field $\varphi$. Moreover, the coupling function $f(\varphi)$ have detractive behavior versus scalar field $\varphi$, too (See Figure \ref{fig3} (B)). As we observe, Figures \ref{fig4} (A) and \ref{fig4} (B) indicate that both scalar potential and coupling function go down versus time which is physically meaningful, for it is well-known that the coupling function and scalar field potential have subtractive behavior versus time. The scalar potential will be constant and the coupling function will vanish in future due to
\begin{equation*}
\lim_{t\rightarrow\infty}V(\varphi)=0.041, \qquad  \qquad \lim_{t\rightarrow\infty}f(\varphi)=0,
\end{equation*}
from (\ref{sol FE}) and (\ref{selections for constants}).\\

Finally, I would like to compare our results with $f(R)$ version of action (\ref{action}) and give a brief comment on Ref. \cite{15}.\\
It is worth to note that unlike our case ($T$-version), the behaviors of the scalar potential and coupling function versus time in $f(R)$ version of the action (\ref{action}) are increasing which is not appreciable (see Ref. \cite{15}). However, some things are uncomforting in that reference; adopted vector field (i.e. $(0;0,0,A(t))$) is not true, because the choice of the last component of vector field being the only nontrivial component violates the cosmological principle on which their metric is based. It seems that their implemented time unit in plotting is 1 Gyr$\equiv \textbf{1}$ (such as our), but one of their scale factors is high at $14 Gyr$ and one is constant that are not admissible. This comparison and discussions which mentioned in Ref. \cite{12} imply that the $T$-version of the action (\ref{action}) which was investigated in this paper and Ref. \cite{12}, works better than $f(R)$ version surveyed in Refs. \cite{15} and \cite{16}.

\section{conclusion}
In this paper, we studied an action in teleparallel gravity, which introduced in Ref. \cite{12}, in the homogeneous and anisotropic (LRS BI) background geometry via the Noether symmetry approach and showed that by the use of cyclic variables we have no consistent solution with the action (\ref{action}), for it produced constant vector field. Then, by keeping Noether currents, we proceeded with B.N.S. approach. We showed that the obtained analytical solution could satisfy Maxwell's equations. Data analysis of the results obtained via this new approach could illustrate exactly the last half of cosmic evolution. These results witnessed that both scale factors have increasing nature and both are close to $1$ at the present time ($14 Gyr$). The amount of deceleration parameter, $q$, is negative throughout this era indicating the accelerating behavior of the universe and is close to minus one at the present time. The resulting model, regarding the amounts of state finders, has a close correspondence with the $\Lambda$CDM universe model. Pursuant to the $Om(z)$ diagnostic, we learned that our universe is in phantom phase. We found that the scalar potential $V(\varphi)$ and coupling function $f(\varphi)$ in this model have detractive behaviors versus time that is physical. Limiting $f(\varphi)$ and $V(\varphi)$ in infinite time give zero and nonzero constants ($\sim 0.041$) respectively, so this model in anisotropic metric predicts that we will have constant scalar potential in the future and $f(\varphi)$ will vanish for its plummeting nature. According to the results of this paper and Ref. \cite{12}, altogether, it seems that $T$-version of the action (\ref{action}) works better than $f(R)$ version. However, in FRW spacetime, both versions of the action (\ref{action}) provide the same field equations.

\end{document}